\documentclass[
aps,
pre,
reprint,
groupedaddress,
]{revtex4-2}
\usepackage{graphicx}
\usepackage{amsmath,amssymb}
\usepackage{bm}
\usepackage{color}
\usepackage[
  colorlinks=true,
  linkcolor=black,
  citecolor=black,
  urlcolor=black,
  bookmarks=true,
  bookmarksnumbered=true
]{hyperref}
\begin{document}

\title{
Transition processes of camphor particle motion in a three-chamber system}

\author{Ryoga Sato}

\affiliation{Department of Physics, Chiba University, Chiba 263-8522, Japan}
\author{Hiroaki Ito}
\affiliation{Department of Physics, Chiba University, Chiba 263-8522, Japan}
\author{Hiroyuki Kitahata}
\email[]{kitahata@chiba-u.jp}
\affiliation{Department of Physics, Chiba University, Chiba 263-8522, Japan}

\begin{abstract}
The motion of a camphor particle in a chamber with three circular regions was investigated through experiments and numerical simulations. Transitions between circular regions were analyzed as discrete stochastic processes.
Experiments revealed the probabilistic transition of the camphor particle between regions, suggesting a tendency for the particle to undergo transitions in the same direction when transitions occur after a short dwell time.
Numerical simulations using a stochastic mathematical model, where the geometry of the experimental system was incorporated through the boundary conditions, reproduced the tendency of transition of the camphor particle between three regions.
\end{abstract}
\maketitle
\section{\label{sec:level1}Introduction}
In non-equilibrium systems, particles can continuously move while gaining and dissipating energy; this behavior is known as self-propelled motion \cite{ramaswamy2010mechanics,vicsek1995novel,ohta2017dynamics,bechinger2016active}. 
Self-propelled motion has attracted interest because it is one of the fundamental characteristics of living organisms \cite{luisi2016emergence,keren2008mechanism,tanimoto2014simple,ariel2014individual}.
However, studying self-propelled motion in biological systems from a physical perspective
is challenging due to environmental fluctuations, inter-individual variability, and the complexity of underlying physiological processes.
To address these challenges, physico-chemical systems such as Janus particles \cite{walther2013janus} and active droplets \cite{masoud2014collective,michelin2023self} have been widely adopted, as they are less complex and more controllable than living organisms, and the mechanisms of their self-propelled motion can be described by mathematical models constructed based on physics and chemistry.
\par 
Achieving control over self-propelled motion may lead to a wide range of applications, such as cargo transport \cite{baraban2012catalytic,simmchen2012asymmetric,sundararajan2008catalytic}, \textit{in vivo} drug delivery \cite{patra2013intelligent}, artificial motors \cite{hong2010biomimetic} and logic gates \cite{gorecki2017xor}. 
Among the various methods for controlling self-propelled motion, the use of boundary geometry provides one potential approach, as self-propelled motion is generally influenced by the boundary shape \cite{zhang2007shape,das2015boundaries,nakata2016selection}. 
For instance, in biological systems, cells migrate through narrow spaces formed by surrounding tissue and the extracellular matrix \cite{charras2014physical, park2015unjamming, friedl2012new}.
As a simplified system for studying self-propelled motion in confined geometries, a two-state system consisting of two regions connected by a narrow bridge has been proposed \cite{bruckner2019stochastic,fink2020area,bruckner2020disentangling,zadeh2024inferring,safara2022model}. This system serves as a model for migration under confined environments encountered by cells in physiological settings, where they migrate through narrow channels or constrictions.
While the two-state system provides an important minimal framework, extending the system to three states can reveal behaviors that do not appear in the two-state system. That is to say, in a three-state system, a particle can select the direction of the transition between regions, allowing for several possible transition modes: transition to the other two regions with an equal probability, transition with a biased probability, or back-and-forth transition between two regions. 
This system provides a simple example of a system with several possible transition modes.\par
As a model of physico-chemical self-propulsion, we have investigated the motion of a camphor particle in confined geometries \cite{nakata2016selection, koyano2016oscillatory,koyano2019rotational,kitahata2022complexity}. Nakata and coauthors investigated a camphor scraping moving between two connected chambers and revealed that its residence in each chamber generates an asymmetric camphor concentration field, which induces its motion toward the other chamber through a surface tension gradient \cite{nakata2000spontaneous}. Motivated by this study, we consider a three-state system using a camphor particle. This approach offers a simple model system that is easy to realize both in experiments and in numerical simulations.
A camphor particle on a water surface is a well-known example of a self-propelled particle interacting through a concentration field \cite{nakata2015physicochemical,nakata1997self,nagayama2004theoretical} as a model system for autodiffusiophoretic motion, which is commonly observed in living organisms \cite{nakajima2014rectified,brown1974temporal}. 
When a camphor particle is placed on water, camphor molecules dissolve from the particle into the surrounding fluid, reducing the surface tension. Fluctuations induce spontaneous symmetry breaking \cite{hayashima2001camphor,koyano2016oscillatory,koyano2019rotational}, which generates a surface tension gradient that drives the particle. Once it starts moving, a higher surface tension in front of the particle sustains its motion through the  sublimation and diffusion of camphor molecules \cite{karasawa2018motion,karasawa2014simultaneous,nomoto2023time}.\par
In this study, we investigate three-state transition processes in a camphor particle system interacting with a concentration field. Specifically, we construct an observation chamber consisting of three circular regions connected by narrow bridges [Fig.~\ref{fig:Experimental setup}(a)], and study the motion of a camphor particle in this setup, analyzing the dynamics from the perspective of three-state transition processes. In addition, we construct a mathematical model that incorporates physico-chemical processes and perform numerical simulations to examine the mechanisms underlying the experimental
observations.
\section{\label{sec:level2}Experimental setup}
\begin{figure}[tb]
\centering
\includegraphics[scale=1.0]{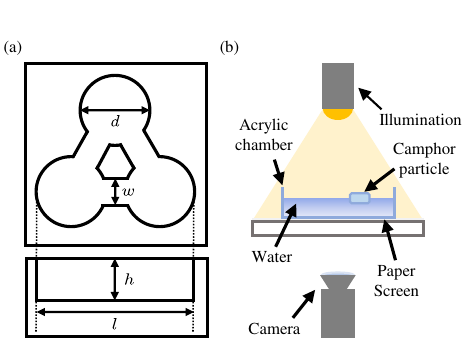}
\caption{Experimental setup.~(a)~Configuration of the three-state chamber.~(b)~Schematic illustration of the experimental system.}
\label{fig:Experimental setup}
\end{figure}

We prepared the chamber by cutting an acrylic plate~(AS ONE Corp., Japan)~into the shape illustrated in Fig.~\ref{fig:Experimental setup}(a)~using a milling machine~(monoFab SRM-20, Roland DG Corp., Japan). The geometrical parameters defined in Fig.~\ref{fig:Experimental setup}(a) were set to $d=31~\text{mm}$, $w=11~\text{mm}$, $h=18~\text{mm}$, and $l=74~\text{mm}$.
(+)-camphor (031-07955, Wako Pure Chemical Industries, Japan) was used without further purification. Pure water was obtained using Elix Essential UV3~(Merck KGaA, Germany). A camphor particle with a diameter of $3~\text{mm}$, a thickness of $1~\text{mm}$ and a mass of $5~\text{mg}$~was prepared using a tablet press~(Kyoto Pastec Co., Ltd., Japan).\par
The schematic illustration of the experimental setup is shown in Fig.~\ref{fig:Experimental setup}(b). 
The prepared particle was placed on the surface of pure water  with a depth of $9~\text{mm}$ in the chamber. The shadow of the camphor particle projected onto a paper screen~(70208, Tamiya, Inc., Japan)~located below the chamber was recorded at $50~\text{fps}$ for $600~\text{s}$ from below using a COMS camera~(STC-MBS43U3V, Omron Sentech Co., Ltd., Japan)~equipped with an objective lens~(VS-LDA25, VS Technology Co., Ltd., Japan). The chamber was illuminated by a white light source (AS3000, AS ONE Corp., Japan) placed above it.
The experiments were performed 25 times under the same conditions. All experiments were conducted at room temperature ($26\pm3~^\circ\text{C}$).
The recorded videos were analyzed using ImageJ \cite{schneider2012nih}.
\begin{figure}[tb]
\centering
\includegraphics[scale=1.0]{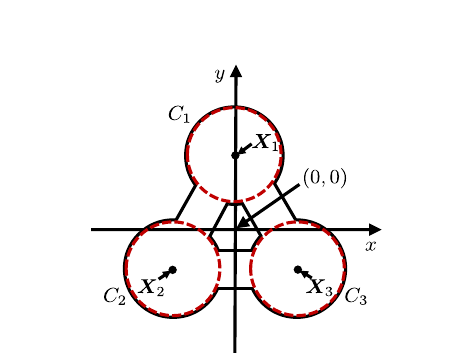}
\caption{Coordinate system used
in this study and the definition of the circular regions $C_{i}$ $(i\in\{1,2,3\})$. The origin coincides with the center of mass (COM) of the triangle formed by $\bm{X}_i$ of the three circular regions $C_{i}$.}
\label{system_coordinate}
\end{figure}
\begin{figure}[tb]
\centering
\includegraphics[scale=1.0]{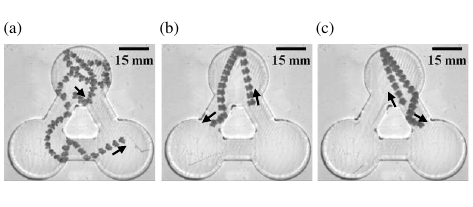}
\caption{Typical behavior of the camphor particle motion observed in the experiments. The black arrow indicates the direction of the motion.
(a) Superposed images of the camphor particle position from $306.60\text{~to~}311.24 
~\text{s}$ with $0.08 
~\text{s}$ intervals. After staying in $C_{1}$, the particle moved to $C_{2}$ via the bridge, and then proceeded toward $C_{3}$.
(b) Superposed images for the two sequential transitions in the same direction from $84.94\text{~to~}85.74 
~\text{s}$ with $0.04 
~\text{s}$ intervals. 
A counterclockwise (CCW) transition from $C_{3}$ to $C_{1}$ was followed by a CCW transition from $C_{1}$ to $C_{2}$.
(c) Superposed images for the two sequential transitions in the different direction from $44.28\text{~to~}45.16 
~\text{s}$ with $0.04 
~\text{s}$ intervals.
A CCW transition from $C_{3}$ to $C_{1}$ was followed by a clockwise (CW) transition from $C_{1}$ to $C_{3}$. 
The camphor particle used in this study was nearly circular; however, the shape of its shadow on the paper screen was distorted by the water surface.
}
\label{exp_trajectory}
\end{figure}
\section{\label{sec:level3}Experimental results}
To describe the experimental results, we introduce a Cartesian coordinate system, whose origin coincides with the center of mass (COM) of the three circular-region centers.
We then define the circular regions $C_i$ whose centers are located at $\bm{X}_i = (X_i, Y_i)$. Here, $i \in \{1,2,3\}$ denotes the region index; $C_1$ lies in the upper half-plane $(Y_1 > 0)$, whereas $C_2$ and $C_3$ lie in the lower half-plane $(Y_2, Y_3 < 0)$ with $X_2 < 0$ and $X_3 > 0$. 
Figure~\ref{system_coordinate} illustrates the coordinate system used in this study and the definition of the circular regions $C_i$.\par
Figure~\ref{exp_trajectory} shows the typical behavior of the camphor particle motion observed in the experiments.
The camphor particle showed self-propelled motion in the three-state chamber. As a typical behavior, the particle stayed in a circular region and then moved to another one through the connecting bridge [Fig.~\ref{exp_trajectory}(a)]. We classified the transitions into two types according to the directions of two sequential transitions: transitions in the same direction [Fig.~\ref{exp_trajectory}(b)] and transitions in the different direction [Fig.~\ref{exp_trajectory}(c)].\par
\begin{figure}[tb]
\centering
\includegraphics[scale=1.0]{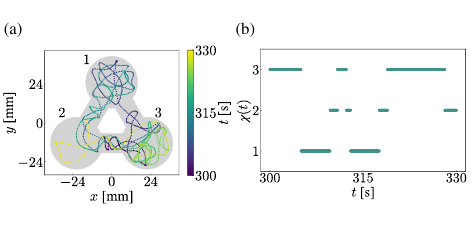}
\caption{Typical experimental results of the camphor particle motion in the three-state chamber, which was obtained within $300$--$330~\text{s}$. (a) Trajectory of the camphor particle. The color denotes the time evolution. (b) Time series of the state $\chi(t)$ corresponding to (a). 
}
\label{experiment_trajectory_chi}
\end{figure}

We obtained the state $\chi(t)$ of the camphor particle at time $t$ from the time series of the COM $\bm r_{\text{c}}(t)$ of the camphor particle. 
Here, we set $t=0$ to the instant at which the camphor particle was placed on the water surface.
The state $\chi$ was defined such that $\chi(t) = i$ when $C_i$ is the most recently visited circular region at time $t$; that is to say, $\chi = i$ from the instant the particle enters $C_i$ until it enters another circular region.
Note that if a particle enters a bridge connecting two circular regions but returns to the original circular region without entering the other circular region, this event is not regarded as a transition.
Figure~\ref{experiment_trajectory_chi} displays the time evolution of the COM $\bm r_{\text{c}}$ of the camphor particle [Fig.~\ref{experiment_trajectory_chi}(a)] and the corresponding time series of the state $\chi$ [Fig.~\ref{experiment_trajectory_chi}(b)]. 
The continuous particle motion in Fig.~\ref{experiment_trajectory_chi}(a) is reduced to a discrete state process in Fig.~\ref{experiment_trajectory_chi}(b), facilitating the analysis of transitions between the circular regions. In Fig.~\ref{experiment_trajectory_chi}(b), the plateaus correspond to intervals during which the most recently visited circular region remains unchanged, whereas jumps indicate transitions to another circular region.
\par
To analyze the behavior of the state transition process, we defined the transition direction $j_{k}$ for the $k$-th transition and the dimensionless dwell time $\tilde\tau_{k}$ before the $k$-th transition using the following equations:
\begin{align}
j_{k} &\equiv 
\chi(t_{k})-\chi(t_{k-1})\quad(\text{mod}~3),\notag\\
&= 
  \begin{cases}
    1, & \text{CCW-direction transition,}\\
    -1, & \text{CW-direction transition,}
  \end{cases}\\
\tilde\tau_{k}&=\dfrac{\bar{v}}{d}(t_{k}-t_{k-1}).
\end{align}
Here, $t_{k}$ represents the time at which the $k$-th transition occurred, $d$ is the diameter of the circular region and $\bar{v}$ is the mean speed of the camphor particle in a single run of the experiment. In our experiments, $d/\bar{v}$ was approximately equal to $0.9~\text{s}$.
We set $t_{0}$ as the starting time of the experiment, i.e., $t_{0}=0$.\par
\begin{figure}[tb]
\centering
\includegraphics[scale=1.0]{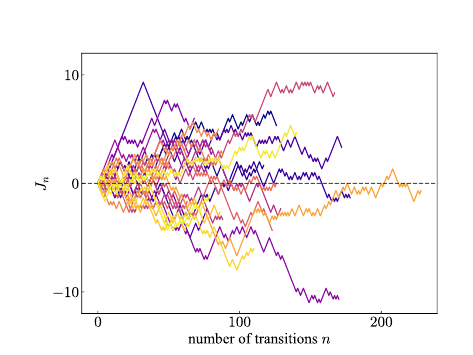}
\caption{Signed cumulated rotation counts $J_{n}$ obtained in 25 independent experiments. 
Each experiment was conducted for $600~\text{s}$, and the number of transitions observed during this period varied among experiments.}
\label{experiment_rotation}
\end{figure}

First, to visualize the transition behavior between the regions, the signed cumulated rotation count $J_{n}$ was defined as
\begin{align}
J_{n} &= \frac{1}{3}\sum_{k=1}^{n}j_{k}.
\label{Signed cumulated rotation number}
\end{align}
The behavior of the signed cumulated rotation counts for all of the 25 independent experiments is shown in Fig.~\ref{experiment_rotation}. 
No consistent preference for clockwise or counterclockwise transitions was apparent
across the runs, although individual runs exhibited fluctuations.\par
Next, to quantitatively characterize the transition behaviors, we defined the transition probabilities for each direction, $P_{\text{ccw}}$ and $P_{\text{cw}}$, as well as the probabilities for two sequential transition, $P_{\text{same}}$ and $P_{\text{diff}}$, as follows:
\begin{align}
P_{\text{ccw}}&=\int_{0}^{\infty}p_{\text{ccw}}(\tilde\tau)\,d\tilde\tau,\\
P_{\text{cw}}&=\int_{0}^{\infty}p_{\text{cw}}(\tilde\tau)\,d\tilde\tau,\\
P_{\text{same}}&=\int_{0}^{\infty}p_{\text{same}}(\tilde\tau)\,d\tilde\tau,\\
P_{\text{diff}}&=\int_{0}^{\infty}p_{\text{diff}}(\tilde\tau)\,d\tilde\tau.
\end{align}
Here, $p_{\text{ccw}}(\tilde\tau),~p_{\text{cw}}(\tilde\tau),~p_{\text{same}}(\tilde\tau)$, and $p_{\text{diff}}(\tilde\tau)$ are the joint probability density functions of the dimensionless dwell time $\tilde\tau$ corresponding to each transition, which are given by
\begin{align}
p_{\text{ccw}}(\tilde\tau)&=p(\tilde\tau_{k}=\tilde\tau,j_{k}
=1),\\
p_{\text{cw}}(\tilde\tau)&=p(\tilde\tau_{k}=\tilde\tau,j_{k}
=-1),\\
p_{\text{same}}(\tilde\tau)&=p(\tilde\tau_{k}=\tilde\tau,j_{k-1}j_{k}
=1),\\
p_{\text{diff}}(\tilde\tau)&=p(\tilde\tau_{k}=\tilde\tau,j_{k-1}j_{k}
=-1).
\end{align}
The probabilities are normalized such that
$P_{\text{ccw}}+P_{\text{cw}}=1,~P_{\text{same}}+P_{\text{diff}}=1$.
The probabilities $P_{\text{ccw}}$, $P_{\text{cw}}$, $P_{\text{same}}$ and $P_{\text{diff}}$ estimated from the experimental data are summarized in Tables \ref{table_of_prob_1} and \ref{table_of_prob_2}. 
These results indicate that two sequential transitions in the same direction are more likely to occur while the probabilities of single transitions are nearly equal.
This suggests that the direction of a previous transition influences the direction of the next transition; i.e., the transition processes of camphor particle motion in a three-chamber system have a memory effect that makes a transition more likely to occur in the same direction as the preceding one.
To confirm the statistical significance of this argument, we evaluated significance by calculating the $p$-value from the $\chi^{2}$ statistic based on a goodness-of-fit test.
The $\chi^{2}$ statistic $Q_{1}$ and $Q_{2}$ for single and two sequential state transitions, respectively, are given by
\begin{table}[tb]
\caption{Transition probability and evaluation of significance for single state transitions obtained from experiments (The total transition count $N_{1}=2727$). Here, $P_{\lambda}~(\lambda=\text{ccw, cw})$ is the observed probability, $\Pi_{\lambda}$ is the theoretical probability, and $Q_{1}$ is the $\chi^{2}$ statistic for single state transitions.}
\begin{center}
\setlength{\tabcolsep}{12pt}
\begin{tabular}{ccccc}
\hline
$P_{\text{ccw}}$&$P_{\text{cw}}$&$\Pi_{\text{ccw}}$&$\Pi_{\text{cw}}$&$Q_{1}$\\ \hline
$0.5020$&$0.4980$&$0.5000$&$0.5000$&$0.044$\\ \hline
\end{tabular}
\end{center}
\label{table_of_prob_1}
\end{table}
\begin{table}[tb]
\caption{Transition probability and evaluation of significance for two sequential state transitions obtained from experiments (The total transition count $N_{2}=2702$). Here, $P_{\lambda}~(\lambda=\text{same, diff})$ is the observed probability, $\Pi_{\lambda}$ is the theoretical probability, and $Q_{2}$ is the $\chi^{2}$ statistic for two sequential state transitions.}
\begin{center}
\setlength{\tabcolsep}{12pt}
\begin{tabular}{ccccc}
\hline
$P_{\text{same}}$&$P_{\text{diff}}$&$\Pi_{\text{same}}$&$\Pi_{\text{diff}}$&$Q_{2}$\\ \hline
$0.5501$&$0.4499$&$0.5000$&$0.5000$&$27.32$\\ \hline
\end{tabular}
\end{center}
\label{table_of_prob_2}
\end{table}
\begin{figure}[tb]
\centering
\includegraphics[scale=1.0]{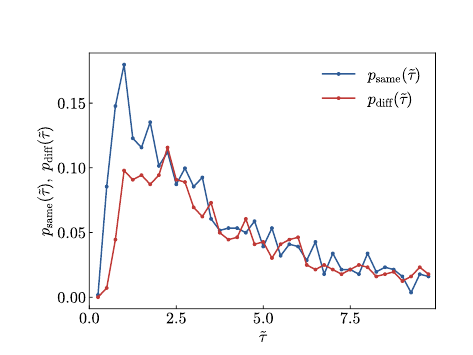}
\caption{Probability density functions depending on the dimensionless dwell time $\tilde{\tau}$ for each sequential transition direction. $p_\text{same}(\tilde{\tau})$ and $p_\text{diff}(\tilde{\tau})$ are shown with blue and red lines.
}
\label{fig:Probability_distribution_exp}
\end{figure}
\begin{align}
Q_{1}&=N_{1}\sum_{\lambda=\text{ccw, cw}}\dfrac{(P_{\lambda}-\Pi_{\lambda})^{2}}{\Pi_{\lambda}},\\
Q_{2}&=N_{2}\sum_{\lambda=\text{same, diff}}\dfrac{(P_{\lambda}-\Pi_{\lambda})^{2}}{\Pi_{\lambda}}.
\end{align}
Here, $P_{\lambda}$ $(\lambda\in\{\text{ccw},~\text{cw},~\text{same},~\text{diff}\})$ is the observed probability and $\Pi_{\lambda}$ is the theoretical probability. $N_{1}$ and $N_{2}$ are the total transition count for single and two
sequential state transitions, respectively.
First, we evaluated the transition probability $P_\text{ccw}$ and $P_\text{cw}$ for single state transitions, which were nearly equal. The theoretical probabilities are $\Pi_\text{ccw}=0.5000$ and $\Pi_\text{cw}=0.5000$. The $\chi^{2}$ statistic obtained from the experiments was $Q_{1}\approx0.044$. The $p$-value was calculated from the $\chi^{2}$ distribution with $1$ degree of freedom, resulting in the $p$-value of approximately $0.83$.
Next, we evaluated the transition probability $P_\text{same}$ and $P_\text{diff}$ for two sequential state transitions to check the memory effect in the transition probability. Assuming that each transition is independent, the theoretical probabilities for two sequential transitions are given by
$\Pi_\text{same}={P_\text{ccw}}^{2}+{P_\text{cw}}^{2}$ and $\Pi_\text{diff}=2P_\text{ccw}P_\text{cw}$.
The obtained theoretical probability were $\Pi_\text{same}=0.5000$ and $\Pi_\text{diff}=0.5000$.
The $\chi^{2}$ statistic obtained from the experimental results was $Q_{2}\approx27.32$, yielding a $p$-value of $<0.01$.
To evaluate the influence of directional bias in each experimental run on the overall results, we define $N_m^{\text{dom}}$ and $N_m^{\text{sub}}$ as the dominant and subordinate numbers of transitions in the CCW and CW directions, respectively, in the $m$-th experimental run.
The transition probability in the dominant and subordinate directions over all experimental runs are then defined as $P_{\text{dom}}=\sum_m N_m^{\text{dom}}/\sum_m(N_m^{\text{dom}}+N_m^{\text{sub}})$ and $P_{\text{sub}}=\sum_m N_m^{\text{sub}}/\sum_m(N_m^{\text{dom}}+N_m^{\text{sub}})$, respectively.
Assuming that each transition is independent, the theoretical probabilities for two
sequential transitions are given by $\tilde{\Pi}_{\text{same}}={P_{\text{dom}}}^2+{P_{\text{sub}}}^2$ and $\tilde{\Pi}_{\text{diff}}=2P_{\text{dom}}P_{\text{sub}}$.
The transition probabilities obtained from the experimental results were $P_\text{dom}=0.5464$ and $P_\text{sub}=0.4536$.
Using these values, the theoretical probabilities were calculated as $\tilde\Pi_\text{same}=0.5043$ and $\tilde\Pi_\text{diff}=0.4957$.
The $\chi^{2}$ statistic obtained from the experimental results was $\tilde Q_{2}\approx22.83$, yielding a $p$-value of $<0.01$.
Based on the above evaluation, we found no statistically significant deviation from equal clockwise and counterclockwise transition probabilities, whereas the probability of two sequential transitions revealed a tendency for the camphor particle to move in the same direction.
The probability density functions $p_{\text{same}}(\tilde\tau)$ and $p_{\text{diff}}(\tilde\tau)$ obtained from the experimental results are shown in Fig.~\ref{fig:Probability_distribution_exp}. Both distributions exhibit a peak at small $\tilde\tau$, but the peak of $p_{\text{same}}(\tilde\tau)$ was higher than that of $p_{\text{diff}}(\tilde\tau)$. As $\tilde\tau$ increases, the discrepancy between the two distributions gradually decreases. 
These results suggest that this system exhibits a memory effect, where the camphor particle tends to undergo sequential transitions in the same direction when the dwell time is short. The characteristic timescale at which the memory effect decays is roughly estimated to be $\tilde\tau\approx2$, i.e. 1.8~s in real dimensions, from Fig.~\ref{fig:Probability_distribution_exp}.
\section{\label{Numerical simulations}Numerical simulations}
We constructed a two-dimensional mathematical model to discuss the motion of the camphor particle in the three-state chamber observed in the experimental system.
We consider the time development of the COM $\bm{r}_{\text{c}}(t)$ of the camphor particle and the camphor concentration field on a water surface $u(t, \bm{x})$ \cite{nagayama2004theoretical}.
\par
The time evolution of $u$ is represented as
\begin{align}
\dfrac{\partial u}{\partial t}&= D_{\text{eff}}\nabla ^{2}u
-a u
+G,
\end{align}
where $D_{\text{eff}}$ is the effective diffusion coefficient \cite{kitahata2018effective,suematsu2014quantitative}, $a$ is the sublimation rate of camphor molecules into the air phase. $G\bigl(\bm{x}, \bm{r}_{\text{c}}(t)\bigr)$ is the
 supply rate of camphor molecules from the camphor particle located at $ \bm{r}_{\text{c}}$, which is given by
\begin{align}
G
=\dfrac{G_{0}}{S_{0}}H_{\epsilon}.
\end{align}
Here, $S_{0}=\pi {R_{\text{c}}}^{2}$ is the surface area of the camphor particle, $R_{\text{c}}$ is the radius of the camphor particle, and $G_{0}$ is the
 total amount of camphor molecules released from the
 camphor particle per unit time. $H_{\epsilon}\bigl(\bm{x}, \bm{r}_{\text{c}}(t)\bigr)$ is a smoothed step function defined as
\begin{align}
H_{\epsilon}=
\frac{1}{2}\left(1+\tanh \frac{R_{\text{c}}-|\bm{x}- \bm{r}_{\text{c}}|}{\epsilon}\right),
\end{align}
where $\epsilon$ is a small positive parameter for smoothing.
\par
The time evolution of $\bm{r}_{\text{c}}$ is given by
\begin{align}
m\dfrac{d^{2}\bm{r}_{\text{c}}}{dt^{2}}&=
-\eta\frac{d\bm{r}_{\text{c}}}{dt}
+\bm{F}_{\text{drv}}
+\bm{F_{\text{rnd}}}
+\bm{F}_{\text{wall}},
\label{eq_of_motion_for_camphor}
\end{align}
where $m$ and $\eta$ are the mass and the resistant coefficient of the camphor particle at the water surface, respectively. $\bm{F}_{\text{drv}}\bigl(\gamma\bigl(u(t,\bm x)\bigr), \bm{r}_{\text{c}}(t)\bigr)$ is the driving force originating from the concentration field, which is described as
\begin{align}
\bm{F}_{\text{drv}}=\iint_{\mathbb{R}^{2}}[\bm\nabla \gamma]H_{\epsilon}\,dS.
\end{align}
Here, $\gamma\bigl(u(t,\bm x)\bigr)$ denotes the surface tension. As the simplest approximation, we assume that $\gamma$ is a linearly decreasing function of the camphor concentration $u$, which is expressed as
\begin{align}
\gamma=\gamma_{0}-\Gamma u,
\end{align}
where $\gamma_{0}$ is the surface tension of pure water and $\Gamma$ is a positive constant. It reflects that the camphor molecules decrease the surface tension of water \cite{karasawa2018motion,suematsu2014quantitative}. $\bm{F_{\text{rnd}}}(t)$ is an effective random force exerted on the camphor, which is represented as
\begin{align}
\bm{F_{\text{rnd}}}&=\sqrt{Y}\bm{\xi},
\label{random_force}
\end{align}
where $Y$ is the intensity of the random force. $\bm{\xi}(t)$ is a stochastic time series that follows the normalized white Gaussian distribution as
\begin{align}
\langle\bm{\xi}(t)\rangle&=\bm{0},\\
\langle\xi_{i}(t)\xi_{j}(s)\rangle&=2\delta_{ij}\delta(t-s).
\end{align}
Here, $\delta_{ij}$ is the Kronecker delta and $\delta(\cdot)$ denotes the Dirac delta function. $\bm{F}_{\text{wall}}\bigl(\bm r_{\text{c}}(t)\bigr)$ is the repulsive force from the wall, which is explicitly modeled as
\begin{align}
\bm{F}_{\text{wall}}
&=-m{\omega_{0}}^{2}\zeta_{\text{c}}\dfrac{\bm{r}_{\text{wall}}}{|\bm{r}_{\text{wall}}|},\\
\zeta_{\text{c}}&=\max\left(0,~R_{\text{c}}+\epsilon-|\bm{r}_{\text{wall}}|\right),\label{F_wall}
\end{align}
where $\bm{r}_{\text{wall}}\bigl(\bm r_{\text{c}}(t)\bigr)$ is the normal vector from $\bm{r}_{\text{c}}$ to the nearest wall. 
For simplicity, we utilize a harmonic potential with an angular frequency $\omega_{0}$ to model wall repulsion.
\par
To simplify the description, we obtain the dimensionless form of the model using $1/a$, $\sqrt{D_{\text{eff}}/a}$, $m$, and $u_{0}$ as the characteristic scales for time, length, mass, and concentration, respectively. The dimensionless variables and coefficients are defined as
\begin{align}
&\tilde t=a t,\quad
\tilde{\bm{x}}=\sqrt{\dfrac{a}{D_{\text{eff}}}}\bm{x}
,
\quad\tilde{\bm{\nabla}}=\sqrt{\dfrac{D_{\text{eff}}}{a}}\bm{\nabla},\notag\\&
\tilde u=\dfrac{1}{u_0}u,\quad
\tilde G_{0}=\dfrac{1}{a S_{0}u_0}G_{0},\quad
\tilde{\bm{r}}_{\text{c}}=\sqrt{\dfrac{a}{D_{\text{eff}}}}\bm{r}_{\text{c}}
,\notag\\&
\tilde\eta=\dfrac{1}{ma}\eta,\quad
\tilde\Gamma=\dfrac{u_{0}}{ma^{2}}\Gamma,\quad
\tilde R_{\text{c}}=\sqrt{\dfrac{a}{D_{\text{eff}}}}R_{\text{c}}
,\notag\\&
\tilde\epsilon=\sqrt{\dfrac{a}{D_{\text{eff}}}}\epsilon,\quad
\tilde Y=\dfrac{1}{m^{2}a^{2}D_{\text{eff}}}Y,\quad
\tilde{\bm{\xi}}=\dfrac{1}{\sqrt{a}}\bm\xi,\notag\\&
\tilde{\bm{r}}_{\text{wall}}=\sqrt{\dfrac{a}{D_{\text{eff}}}}\bm{r}_{\text{wall}},\quad
\tilde \omega_{0}=\dfrac{1}{a}\omega_{0}.
\end{align}
\begin{figure*}[tb]
\centering
\includegraphics[scale=1]{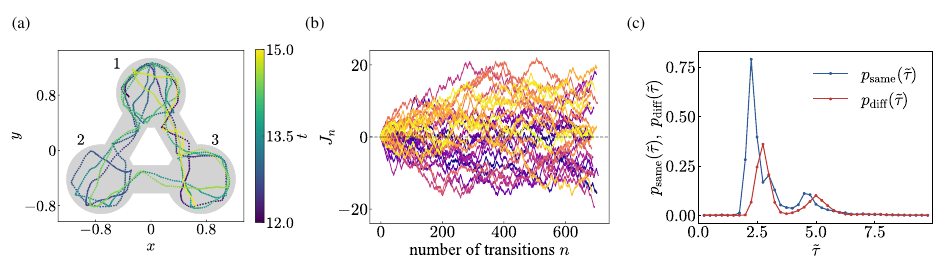}
\caption{Results of numerical simulations using the mathematical model of the camphor particle [Eqs.~(\ref{dimensionless_reaction_diffusion_eq})--(\ref{zeta})].
(a) Typical trajectory of a camphor particle.
(b) Signed cumulated rotation counts.
(c) Probability density functions depending on the dimensionless dwell time for each sequential transition direction.
Considering the relationship between the experimental and numerical simulation scales, one unit of simulation time corresponds to 10 s in the actual experiments. Therefore, the trajectory in (a) corresponds to 30 s in the actual experiment, and the total simulation duration is 180, which is three times as long as the experimental duration.
}
\label{fig:simulation_plot}
\end{figure*}
Hereafter, we omit $\tilde\cdot$ for simplicity. Dimensionless forms of the model are summarized as
\setlength{\jot}{4pt} 
\begin{gather}
\dfrac{\partial u}{\partial t}
= \nabla^{2}u
 - u
 + G_{0}H_{\epsilon},\label{dimensionless_reaction_diffusion_eq}\\
\dfrac{d^{2}\bm{r}_{\text{c}}}{dt^{2}}
= -\eta\dfrac{d\bm{r}_{\text{c}}}{dt}
 - \Gamma\iint_{\mathbb{R}^{2}}[\bm{\nabla}u]
   H_{\epsilon}\,dS
 + \bm{F}_{\text{rnd}}
 + \bm{F}_{\text{wall}}, \label{dimensionless_eq_of_motion}\\
H_{\epsilon}
= \frac{1}{2}\left(1+\tanh\frac{R_{\text{c}}-|\bm{x}-\bm{r}_{\text{c}}|}{\epsilon}\right), \\
\bm{F}_{\text{rnd}}
= \sqrt{Y}\,\bm{\xi}, \label{noise}\\
\langle\bm{\xi}(t)\rangle
= \bm{0}, \\
\langle\xi_{i}(t)\xi_{j}(s)\rangle
= 2\delta_{ij}\delta(t-s), \\
\bm{F}_{\text{wall}}
=-m{\omega_{0}}^{2}\zeta_{\text{c}}\dfrac{\bm{r}_{\text{wall}}}{|\bm{r}_{\text{wall}}|},\\
\zeta_{\text{c}}=\max\left(0,~R_{\text{c}}+\epsilon-|\bm{r}_{\text{wall}}|\right)\label{zeta}.
\end{gather}
To ensure the three-fold rotational symmetry in the numerical simulations, a triangular lattice was used for the spatial discretization of the concentration field (details are shown in Appendix \ref{AppendixA}).
The numerical simulations were performed using the Euler method for Eq.~(\ref{dimensionless_reaction_diffusion_eq}) and the Euler-Maruyama method \cite{higham2001algorithmic} for Eq.~(\ref{dimensionless_eq_of_motion}). The geometrical parameters defined in Fig.~\ref{fig:Experimental setup}(a) were set to $d=1$,
$w=0.36$ and $l=2.44$ in the numerical simulations.
In the same manner as in the experiments, we defined $C_i$ as the circular regions whose centers are at $\bm{X}_i = (X_i, Y_i)$ ($i \in \{1,2,3\}$).
We adopted the Neumann boundary condition for $u$ at the periphery of the chamber. 
We imposed a zero diffusive flux across the
chamber boundary.
The time step and spatial mesh were set to be $\Delta t=0.00001$ and $\Delta l=0.02$, respectively.
The parameters used in the calculations were $G_{0}={400}/{\pi}$, $R_{\text{c}}=0.05$,
$\epsilon=0.02$,
$\eta=200$, $\Gamma=240000$ and $Y=1000$.
With regard to the parameters describing the repulsive force from the wall, we set
$\omega_{0}={\pi}/{0.002}$, so that the effective collision time with the wall was $\tau_{\text{wall}}=0.002$. We performed 32~numerical simulations with different initial conditions using the same set of parameters.
The initial position of the camphor particle was set to $\bm{X}_{1} + (\Delta x, \Delta y)$, where $(\Delta x, \Delta y)=(0,\sqrt{3/100})$. The initial velocity of the camphor particle was set to $|v_{0}|=0.1$, while the direction was sampled as $\theta_{0,j}=0.001+j\Delta\theta~(j\in\{0, 1,...,31\})$, with $\Delta\theta=(2\pi-0.003)/31$. We adopted the Box-Muller method to generate the normalized white Gaussian noise, and different random seeds were used for noise generation in all numerical simulations. The concentration field was initially zero at every point in the region.
\section{\label{sec:level5}Simulation results}
Figures~\ref{fig:simulation_plot}(a)--(c) show the time evolution of the COM $\bm r_{\text{c}}$ of the camphor particle [Fig.~\ref{fig:simulation_plot}(a)], the cumulated rotation counts [Fig.~\ref{fig:simulation_plot}(b)], and the joint probability density function depending on the dimensionless dwell time [Fig.~\ref{fig:simulation_plot}(c)] obtained from the numerical simulation with a random force ($Y=1000$).
As in the experimental results, both distributions exhibit a peak at small $\tilde\tau$, where the peak of $p_{\text{same}}(\tilde\tau)$ is higher than that of $p_{\text{diff}}(\tilde\tau)$, and the discrepancy between the two distributions gradually decreases as $\tilde\tau$ increases.
As in the experiments, we calculated $P_{\text{same}}$ and $P_{\text{diff}}$ from the simulation data, resulting in $P_{\text{same}}\approx0.64$ and $P_{\text{diff}}\approx0.36$.
The characteristic decay time scale of the memory effect was roughly estimated to be $\tilde\tau\approx2.5$ from Fig.~\ref{fig:simulation_plot}(c). 
\section{\label{sec:level6}Discussion}
First, we discuss the quantitative aspects of numerical simulation.
For the characteristic time and characteristic length, we set $a=0.1\text{~s}^{-1}$ and $\sqrt{{D_{\text{eff}}}/{a}}=30\text{~mm}$. These values are comparable to those reported in a previous experimental study \cite{suematsu2014quantitative}. The parameters used in the numerical simulations were set to $R_{\text{c}}=1.5\text{~mm}$,
$d=30\text{~mm}$, $w=10.8\text{~mm}$, $l=73.2\text{~mm}$. In the experiments, $R_{\text{c}}=1.5\text{~mm}$,
$d=31\text{~mm}$, $w=11\text{~mm}$, and $l=74\text{~mm}$. Therefore, these values in the numerical
simulations are close to their experimental
counterparts.
The mass of the camphor particle in the experiments was $m=5\text{~mg}$, and the ratio of the mass to the viscous resistance coefficient was estimated in a previous study to be $m/\eta=5.0\times10^{-2}\text{~s}$ \cite{koyano2019rotational}.
Thus, the dimensionless resistance coefficient $\tilde\eta=\eta/(ma)=200$ was adopted in the numerical simulations. The dimensionless interfacial tension coefficient $\tilde\Gamma$ was set such that the dimensionless steady-state velocity $\tilde v_{\text{st}}$ of the camphor particle was $\tilde v_{\text{st}}\approx15$. The actual steady-state velocity $v_{\text{st}}$ is estimated to be $v_{\text{st}}=\tilde v_{\text{st}}\sqrt{{aD_{\text{eff}}}}\approx45~\text{mm/s}$, which agrees in order of magnitude with the velocity of camphor particles observed experimentally. Estimating the magnitude of the driving force in the steady state yields $F=\eta v_{\text{st}}\approx10^{-6}\text{ N}$, which also agrees with the values reported in the previous study \cite{suematsu2014quantitative}. 
In our numerical simulations, $d/\bar{v}$ was approximately equal to $0.8~\text{s}$ in real dimensions, which is reasonably close to the experimental results.
Thus, we conclude the parameters used in the numerical simulations correspond to the experimental conditions.
\par
Next, we discuss the origin and validity of the noise introduced in the numerical simulations. 
In the experiments, the motion of the camphor particle can be influenced by various effects that may be regarded as noise, such as hydrodynamic interactions and inevitable slight shape anisotropy.
In the numerical simulations without noise ($Y=0$), the trajectory exhibits a limit cycle with repeated transitions in the same direction, unlike the experimental behavior.
Therefore, these effects are collectively represented by a stochastic term in the numerical simulations, with a dimensionless noise intensity of $Y=1000$. Consequently, in the numerical simulations, the camphor particle moved at $v\approx10$ and reached a maximum velocity of $v_{\text{max}}\approx30$. Converting these to real dimensions yields $v\approx30\text{~mm/s}$ and $v_{\text{max}}\approx90\text{~mm/s}$, which approximately agreed with experimental results. This discussion indicates that the chosen noise intensity value is reasonable to reflect the present experimental conditions.
\par
Finally, we discuss the mechanism of the state transition processes of camphor particle motion observed in this study.
The camphor particle tended to undergo transitions in the same direction when the dimensionless dwell time was short, and this tendency disappeared as the dimensionless dwell time increased. 
A possible explanation is that elevated camphor concentration persists along the
recent trajectory of the particle, suppressing back transitions for short dwell times. With increasing dwell time, sublimation gradually reduces the residual camphor concentration and weakens this memory effect, leading to nearly equal transition probabilities.
This can be understood as a memory effect \cite{dos1995free,jackson2004trail,hughes2004value} arising from the concentration field \cite{nakata2013motion, nakata2018reciprocating,koyano2020simple}.
In the present study, the duration of the memory effect observed in the experiments was $\tau\approx1.8\text{~s}$, which is comparable to the value obtained in the numerical simulations, $\tilde\tau\approx2.5$, corresponding to $\tau\approx2.0\text{~s}$ in real dimensions. \par
To examine the relationship between the state-transition processes and the concentration field observed in the experiments and numerical simulations, we performed numerical simulations using a mathematical model in which the driving force term arising from the concentration field was replaced with a constant self-propulsion force independent of the concentration field. This model failed to reproduce the experimental results (see Appendix \ref{AppendixB} for details). This suggests that the concentration field retains information about the transition direction and that the memory effect of the concentration field is essential to the observed transition processes.
\par
In the present study, we adopted only a chamber geometry illustrated in Fig.~\ref{fig:Experimental setup}(a). Modifying the geometry, for example by changing the bridge width or the shape of the regions (e.g., to a ratchet shape), is expected to give rise to qualitatively different transition dynamics. Investigating such geometric effects remains a topic for future work.
\section{\label{sec:level7}Conclusions}
In the present study, we investigated the motion of a camphor particle in a chamber connecting three circular regions through experiments and numerical simulations, to analyze the transitions as discrete stochastic processes. The experiments revealed a tendency for the particle to undergo two sequential transitions in the same direction after short dwell times. A mathematical model consisting of an equation of motion for a camphor particle with an effective stochastic term and a reaction-diffusion equation for the camphor concentration field with experimental geometry as boundary conditions qualitatively reproduced the observed transition processes. 
These results support the possibility that the observed transitions are associated with a memory effect of the concentration field, which suppresses backward transitions at short dwell times and weakens with increasing dwell time.
This study suggests that a memory of the particle motion encoded in the concentration field plays an important role in the state-transition processes, providing insight into the dynamics of self-propelled motion with concentration-field-induced memory effects.

\begin{acknowledgments}
We are grateful to Prof. Natsuhiko Yoshinaga for valuable discussions.
This work was supported by JSPS KAKENHI Grants Nos.~JP24K06978, JP24K22311, JP24K06972, JP25K00918, JP26K00671, MEXT KAKENHI Grant No.~JP26H00384, JSPS Bilateral Program Number JSPSBP120264601, and the Cooperative Research Program of ``Network Joint Research Center for Materials and Devices'' (No.~20264003). This work was also supported by the JSPS Core-to-Core Program ``Advanced core-to-core network for the physics of self-organizing active matter (JPJSCCA20230002)'' and by MEXT Promotion of Distinctive Joint Usage/Research Center Support Program Grant Number JPMXP0724020292.
\end{acknowledgments}
\appendix
\section{Discretization of the Laplacian and gradient
on a triangular lattice\label{AppendixA}}
\begin{figure}[tb]
\centering
\includegraphics[scale=1.0]{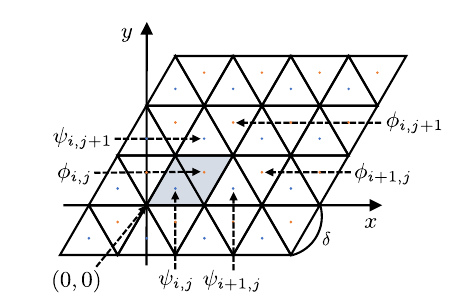}
\caption{Schematic illustration of the triangular lattice employed in the present work.}
\label{triangle_grid}
\end{figure}

The system geometry in this experiment possesses three-fold rotational symmetry. However, numerical calculations based on the partial differential equations using a standard square lattice may break this symmetry. To preserve the three-fold rotational symmetry in numerical simulation, a triangular lattice was employed for the spatial discretization. First, the domain is divided into a triangular lattice with a side length of $\delta$ as shown in Fig.~\ref{triangle_grid} and field variables $\psi$ and $\phi$ are introduced to the
centroids of the triangular lattice. Here, $\psi$ represents the field variables on the triangular lattice pointing in the positive $y$-direction, while $\phi$ represents that on the triangular lattice pointing in the negative $y$-direction. Next, we derive the expression for the discretization of the Laplacian on the triangular lattice. Integrating the continuous equation and using Fick's law $\bm{J}=-D\bm{\nabla}\psi$ yields the following relationship:
\begin{align}
\int_{\Omega}\displaystyle\frac{\partial \psi}{\partial t}\;dS&=-\int_{\Omega}\bm{\nabla}\cdot{\bm{J}}\;dS\notag\\
&=-\oint_{\partial \Omega}\bm{J}\cdot{\bm{n}}\;dl\notag\\
&=D\oint_{\partial \Omega}[\bm{\nabla}\psi]\cdot{\bm{n}}\;dl.
\end{align}
By discretizing the above equation, we obtain the representation of $\psi^{n+1}_{i,j}$ at the $(n+1)$-th step:
\begin{align}
\psi^{n+1}_{i,j}&=\psi^{n}_{i,j}
+\displaystyle\frac{12D\Delta t}{\delta^{2}}
\left[
\frac{\phi^{n}_{i-1,j}+\phi^{n}_{i,j}+\phi^{n}_{i,j-1}}{3}-\psi^{n}_{i,j}
\right].
\end{align}
Here, we use $\Delta S=\sqrt{3}\delta^{2}/4$, $\Delta l=\delta$, and the fact that the distance between adjacent $\psi$ and $\phi$ is $\delta/\sqrt{3}$. Similarly, for $\phi^{n+1}_{i,j}$, we have
\begin{align}
\phi^{n+1}_{i,j}&=\phi^{n}_{i,j}
+\displaystyle\frac{12D\Delta t}{\delta^{2}}
\left[
\frac{\psi^{n}_{i,j}+\psi^{n}_{i+1,j}+\psi^{n}_{i,j+1}}{3}-\phi^{n}_{i,j}
\right].
\end{align}
Through the above procedures, we obtain an expression for the discretization of the Laplacian on a triangular lattice.\par
Next, we derive an expression for the discretization of the gradient vector on a triangular lattice. The following approximation holds for the gradient:
\begin{align}
\psi(\bm{r}+\Delta\bm{r})\approx \psi(\bm{r})+[\bm{\nabla}\psi(\bm{r})]\cdot{\Delta{\bm{r}}}.
\end{align}
Discretizing the above equation with $\Delta \bm {r}=\delta\bm e(\varphi)/\sqrt{3}$ for $\varphi=\pi/6,~5\pi/6,~\text{and}~3\pi/2$, where $\bm e(\varphi)=\cos\varphi\;\bm e_{x}+\sin\varphi\;\bm e_{y}$, yields the following equations:
\begin{align}
\phi^{n}_{i,j}&=\psi^{n}_{i,j}+\dfrac{\delta}{\sqrt{3}}\left[\bm{\nabla}\psi\right]^{n}_{i,j}\cdot\bm{e}\left(\dfrac{\pi}{6}\right),\\
\phi^{n}_{i-1,j}&=\psi^{n}_{i,j}+\dfrac{\delta}{\sqrt{3}}\left[\bm{\nabla}\psi\right]^{n}_{i,j}\cdot\bm{e}\left(\dfrac{5\pi}{6}\right),\\
\phi^{n}_{i,j-1}&=\psi^{n}_{i,j}+\dfrac{\delta}{\sqrt{3}}\left[\bm{\nabla}\psi\right]^{n}_{i,j}\cdot\bm{e}\left(\dfrac{3\pi}{2}\right).
\end{align}
Here, $\left[\bm{\nabla}\psi\right]^{n}_{i,j}$ is the discretized gradient vector of the concentration field $\psi^{n}_{i,j}$ at time step $n$, and for each directional component,
$\left[\bm{\nabla}\psi\right]^{n}_{i,j}=\left[(\bm{\nabla}\psi)_{x}\right]^{n}_{i,j}\bm{e}_{x}+\left[(\bm{\nabla}\psi)_{y}\right]^{n}_{i,j}\bm{e}_{y}$.
Using these, we obtain the following expression:
\begin{align}
\left[(\bm{\nabla}\psi)_{x}\right]^{n}_{i,j}&=\displaystyle\frac{\phi^{n}_{i,j}-{\phi}^{n}_{i-1,j}}{\delta},\\
\left[(\bm{\nabla}\psi)_{y}\right]^{n}_{i,j}&=\displaystyle\frac{{\phi}^{n}_{i,j}+{\phi}^{n}_{i-1,j}-2\phi^{n}_{i,j-1}}{\sqrt{3}\delta}.
\end{align}
Likewise for $\left[\bm{\nabla}\phi\right]^{n}_{i,j}$, we obtain
\begin{align}
\left[(\bm{\nabla}\phi)_{x}\right]^{n}_{i,j}&=\displaystyle\frac{\psi^{n}_{i+1,j}-{\psi}^{n}_{i,j}}{\delta},\\
\left[(\bm{\nabla}\phi)_{y}\right]^{n}_{i,j}&=\displaystyle\frac{2\psi^{n}_{i,j+1}-{\psi}^{n}_{i,j}-{\psi}^{n}_{i+1,j}}{\sqrt{3}\delta}.
\end{align}
Through the above procedure, we derive the expression for the discretization of gradient vectors on a triangular lattice to calculate $\bm F_{\text{drv}}$.
\begin{figure*}[tb]
\centering
\includegraphics[scale=1]{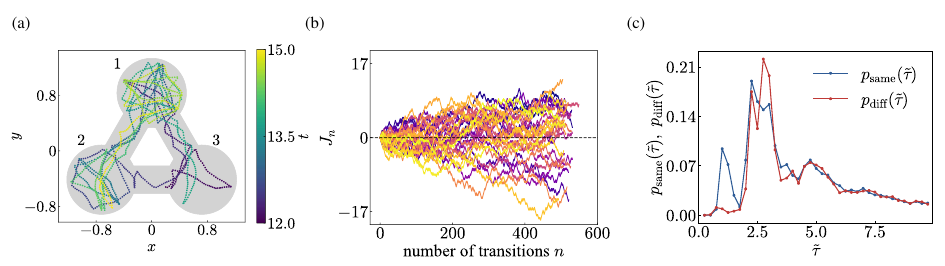}
\caption{Results of numerical simulations using a mathematical model of the self-propelled particle with a constant driving force [Eqs.~(\ref{noise})--(\ref{zeta}), (\ref{SPP-v})--(\ref{eq_of_motion_for_ABP})]. (a) Typical trajectory of a self-propelled particle. (b) Signed cumulated rotation counts. (c) Probability density functions depending on the dimensionless dwell time for each sequential transition direction.
Considering the relationship between the experimental and numerical simulation scales, one unit of simulation time corresponds to 10 s in the actual experiments. Therefore, the trajectory in (a) corresponds to 30 s in the actual experiment, and the total simulation duration is three times as long as the experimental duration.
}
\label{simulation_abp_plot}
\end{figure*}
\section{
Numerical simulations for a self-propelled particle with a constant driving force
\label{AppendixB}}
We consider a simplified model in which the driving force term $\bm{F}_{\text{drv}}$ arising from the concentration field in Eq.~(\ref{eq_of_motion_for_camphor}) in the main text is replaced by a constant driving force $\hat{\bm{F}}_{\text{drv}}$ acting in the direction of the particle velocity and independent of the concentration field $u$ \cite{ishikawa2025simple}. Namely, the driving force is represented as
\begin{align}
\hat{\bm{F}}_{\text{drv}}&=f\frac{\bm{v}_{\text{c}}}{|\bm{v}_{\text{c}}|}.
\end{align}
Here, $\bm{v}_{\text{c}}={d\bm{r}_{\text{c}}}/{dt} $. The equation of motion for a self-propelled particle with a constant driving force is obtained as follows:
\begin{align}
\dfrac{d\bm{r}_{\text{c}}}{dt} &= \bm{v}_{\text{c}},\\
m\dfrac{d\bm{v}_{\text{c}}}{dt} &= 
-\eta\bm{v}_{\text{c}}
+\hat{\bm{F}}_{\text{drv}}
 + \bm{F}_{\text{rnd}}
 + \bm{F}_{\text{wall}}.
\end{align}
Note that the definitions of $\bm{F}_{\text{rnd}}$ and $\bm{F}_{\text{wall}}$ are identical to Eqs.~(\ref{random_force})--(\ref{F_wall}) in Sect.~\ref{Numerical simulations} in the main text.
To simplify the description, we obtain the dimensionless form of the model using ${1}/{a}$, $\sqrt{{D_{\text{eff}}}/{a}}$, and $m$ as the characteristic scales for time, length, and mass, respectively.
The dimensionless variables and coefficients are defined as
\begin{align}
\tilde{\bm{v}}_{\text{c}}=\dfrac{1}{\sqrt{{aD_{\text{eff}}}}}\bm{v}_{\text{c}},\quad
\tilde f=\dfrac{1}{m\sqrt{{a^{3}D_{\text{eff}}}}}f.
\end{align}
Hereafter, we omit $\tilde\cdot$ for simplicity. Dimensionless forms are summarized as
\begin{gather}
\dfrac{d\bm{r}_{\text{c}}}{dt} = \bm{v}_{\text{c}},\label{SPP-v}\\
\dfrac{d\bm{v}_{\text{c}}}{dt} = 
-\eta\bm{v}_{\text{c}}
+\hat{\bm{F}}_{\text{drv}}
 + \bm{F}_{\text{rnd}}
 + \bm{F}_{\text{wall}},\\
\hat{\bm{F}}_{\text{drv}}=f\frac{\bm{v}_{\text{c}}}{|\bm{v}_{\text{c}}|}.\label{eq_of_motion_for_ABP}
\end{gather}
where $\bm{F}_{\text{rnd}}$ and $\bm{F}_{\text{wall}}$ are the same as those given in Eqs.~(\ref{noise})--(\ref{zeta}) in Sect.~\ref{Numerical simulations} in the main text.
The numerical calculation was performed using the Euler-Maruyama method \cite{higham2001algorithmic} for Eqs.~(\ref{SPP-v})--(\ref{eq_of_motion_for_ABP}). The parameter $f$ was set to $f=3000$, while the other parameters and initial conditions were identical to those used in the simulations in Sect.~\ref{Numerical simulations} in the main text.
Note that the particle speed never reached zero in these numerical simulations.
Here, estimating the magnitude of the driving force at a steady state yields $f=\tilde fm\sqrt{{a^{3}D_{\text{eff}}}}\approx10^{-6}~\text{N}$, which agrees with the value reported in the previous study \cite{suematsu2014quantitative}.\par
Figures~\ref{simulation_abp_plot}(a)--(c) show the time evolution of the COM $\bm r_{\text{c}}$ of the self-propelled particle with a constant driving force [Fig.~\ref{simulation_abp_plot}(a)], the cumulated rotation counts [Fig.~\ref{simulation_abp_plot}(b)], and the joint probability density function depending on the dimensionless dwell time [Fig.~\ref{simulation_abp_plot}(c)] obtained from the numerical simulation with a random force ($Y=1000$). 
In our numerical simulations, $d/\bar{v}$ was approximately equal to $0.7~\text{s}$ in real dimensions, which is reasonably close to the experimental results.
The behavior of the joint probability density function of the dimensionless dwell time showed a maximum peak at $\tilde\tau\approx3$ in $p_{\text{diff}}(\tilde\tau)$, whereas the maximum peak was observed in $p_{\text{same}}(\tilde\tau)$ in the experiments. Thus, numerical simulations based on the mathematical model, in which the driving force from the concentration field was replaced by a constant driving force, could not reproduce the experimental results. This suggests that the memory of the concentration field, which reflects the past transition directions of the camphor particle, plays a significant role in the state-transition processes.
\bibliography{references}
\end{document}